\documentclass[11pt]{article}
\usepackage{latexsym}

\usepackage{epsfig}
\usepackage{graphicx}

\catcode`\@=11

\@addtoreset{equation}{section}

\def\@normalsize{\@setsize\normalsize{15pt}\xiipt\@xiipt
\abovedisplayskip 14pt plus3pt minus3pt%
\belowdisplayskip \abovedisplayskip
\abovedisplayshortskip  \z@ plus3pt%
\belowdisplayshortskip  7pt plus3.5pt minus0pt}
\def\small{\@setsize\small{13.6pt}\xipt\@xipt
\abovedisplayskip 13pt plus3pt minus3pt%
\belowdisplayskip \abovedisplayskip
\abovedisplayshortskip  \z@ plus3pt%
\belowdisplayshortskip  7pt plus3.5pt minus0pt
\def\@listi{\parsep 4.5pt plus 2pt minus 1pt
            \itemsep \parsep
            \topsep 9pt plus 3pt minus 3pt}}

\def\underline#1{\relax\ifmmode\@@underline#1\else
        $\@@underline{\hbox{#1}}$\relax\fi}
\@twosidetrue \relax

\catcode`@=12

\catcode`\@=11

\def\section{\@startsection{section}{1}{\z@}{3.5ex plus 1ex minus
   .2ex}{2.3ex plus .2ex}{\large\bf}}

\def\ps@headings{\def\@oddfoot{}\def\@evenfoot{}
\def\@oddhead{\hbox{}\hfill
        \makebox[.5\textwidth]{\raggedright\ignorespaces --\thepage{}--
        \hfill }}
\def\@evenhead{\@oddhead}
\def\subsectionmark##1{\markboth{##1}{}}
}

\ps@headings

\catcode`\@=12

\newcommand{\be}{\begin{equation}}
\newcommand{\ee}{\end{equation}}
\newcommand{\bea}{\begin{eqnarray}}
\newcommand{\nn}{\nonumber}
\newcommand{\eea}{\end{eqnarray}}

\begin{document}

\begin{titlepage}
\begin{flushright}
hep-th/0601152\\ January 2006
\end{flushright}

\vspace{0.30in}
\begin{centering}

{\large {\bf Geometrical Tachyon Dynamics in the Background of a
Bulk Tachyon Field }}
\\

\vspace{0.7in} {\bf  Eleftherios Papantonopoulos$^{~\alpha,*}$,
Ioanna Pappa$^{~\alpha, \beta,\dag}$ and Vassilios
Zamarias$^{~\alpha,\flat}$}

\vspace{0.04in}

$^{\alpha}$Department of Physics, National Technical University of Athens,\\
Zografou Campus GR 157 73, Athens, Greece\\
$^{\beta}$Department of Physics, Stockholm University, Stockholm
SE 106 91, Sweden\\
\end{centering}

\vspace{0.8in}

\begin{abstract}
 We study the dynamics of a
D3-brane moving in the background of a bulk tachyon field of a
D3-brane solution of Type-0 string theory. We show that the
dynamics on the probe D3-brane can be described by a geometrical
tachyon field rolling down its potential which is modified by a
function of the bulk tachyon and inflation occurs at weak string
coupling, where the bulk tachyon condenses, near the top of the
geometrical tachyon potential. We also find a late accelerating
phase when the bulk tachyon asymptotes to zero which in the
geometrical tachyon picture corresponds to the minimum of the
geometrical potential.

 \end{abstract}

\begin{flushleft}

 \vspace{0.7in} $^{*}$lpapa@central.ntua.gr \\
 $^{\dag}$ gianna@physto.se\\
$^{\flat}$zamarias@central.ntua.gr
\end{flushleft}
\end{titlepage}

\section{Introduction}

In open string theory the presence of a tachyon field indicates an
instability in its world-volume. There is strong evidence of a
relation between the full dynamical evolution in string theory and
renormalization group flows on the world-sheet. The two sides have
many features in common. The most profound one is that a
world-sheet RG flow away from an unstable string background ends
at an infrared conformal field theory that may generically be
expected to be stable. Similarly, the dynamical process of tachyon
condensation is generically expected to decay into a stable
solution of string theory \footnote{For reviews on open tachyon
condensation see \cite{Taylor:2002uv,Sen:2004nf}.}.

The time evolution of a decaying D-brane in an open string theory
can be described by an exact solution called rolling tachyon
\cite{Sen:2002nu,Sen:2002in} or S-brane
\cite{Gutperle:2002ai,Maloney:2003ck}. The homogeneous decay can
be described by perturbing the D-brane boundary conformal field
theory. During this decay described by an instability of the RG
flow on the world-sheet of the string, the spacetime energy
decreases along the RG flow. The end point of this evolution is to
dump the energy released by the tachyon condensation into the
surrounding space, presumably in the form of closed string
radiation, and then relax to its ground state.

Closed string theories are theories of gravity and spacetime is
dynamical in such theories. An instability in the spacetime theory
implies also an instability of RG flow, on the world-sheet of the
string. Therefore we expect similar behaviour as in the open
string case. There are however some important differences. The
condensation of a closed tachyon field modifies the asymptotics of
spacetime (for a review on closed tachyon condensation see
\cite{Headrick:2004hz}). Nevertheless it was found that the
spacetime energy decreases along bulk world-sheet RG flows, at
least for the flows for which this statement may be sensibly
formulated. Also, for the case of the closed tachyons,
conservation of energy severely complicates the issue of whether
condensation leads to the true vacuum of the theory, if it has
one.

The Dirac-Born-Infeld action as an effective theory successfully
describes the physics in the world-volume of string theory. In the
case of open string theory, the DBI action was employed to
describe the dynamics of an open tachyon field and in particular
the time evolution of a D-brane by the rolling of a tachyon field
down its potential. When a closed tachyon field is present in the
bulk, the world-volume dynamics is described by a modified DBI
action. In the case of Type-0 string theory, the world-volume
couplings of the tachyon with itself and with massless fields on a
D-brane were calculated \cite{Garousi:1998fg,Garousi:1999fu}. It
was found that the bulk tachyon appears as an overall coupling
function in the DBI action \cite{Garousi:1998fg,Klebanov:1998yy}.

Type-0 string theories~\cite{Polyakov:1998ju} are interesting
because of their connection
 \cite{Klebanov:1998yy,Klebanov1:1998yy} to four-dimensional
 $SU(N)$ gauge theory. These string theories do not have
spacetime supersymmetry and as a result of GSO projection a closed
tachyon field appears in their action. The tachyon field in the
action appears with its potential and a function $f(T_{bulk})$
which couples to the RR flux of the background. It was shown
\cite{Klebanov:1998yy,Klebanov1:1998yy} that if this function has
an extremum then, because of its coupling to the RR field, it
stabilizes the closed tachyon potential driving its mass to
positive values \cite{Klebanov:1999um}. Type 0B string theories
have also in general open string tachyons in their
spectra~\cite{Bianchi:1990yu}. However, it can be shown that large
N gauge theories, which are constructed on N coincident D-branes
of Type-0  theory contain no open
tachyons~\cite{Klebanov:1998yy,Polyakov:1998ju}

Another appealing feature of Type-0 string theories is that, when
the tachyon field is non trivial, because of its coupling to
dilaton field, there is a renormalization group flow from infrared
(IR) to ultra violet (UV) \cite{Minahan:1998tm,Bigazzi:2001ta}.
This corresponds to a flow of the couplings of the dual gauge
theory from strong (IR region) to weak couplings (UV region) where
the dilaton field gets small and the tachyon field receives a
constant value. However, because the evolution equations are
complicated when the tachyon and dilaton fields are not constant,
the analytic evolution from IR to UV fixed points is not known,
and only the behaviour of the theory near the fixed points is well
understood. At these two fixed points it has been shown that the
background geometry asymptotes to the near-horizon $AdS_{5}\times
S^{5}$ geometry.

The Type-0 string theory is an example of a closed string theory,
where the tachyon condensation stabilizes the theory. We are far
from understanding its full dynamics but nevertheless it gives us
some information about the gravitational dynamics of the bulk. It
would be interesting to see what is the effect of the closed
tachyon condensation on the boundary theory.

In this work we will consider a probe D3-brane moving in the
background of a Type-0 string. We will study a particular D3-brane
bulk solution of this string theory, for which we know an exact
solution at least in the weak string limit. The probe D3-brane
will be affected by the bulk geometry through a modified DBI
action describing its dynamics, and through the Wess-Zumino term
which encodes the structure of the bulk. In spite that there is no
any open tachyon on the moving probe D3-brane, the movement of the
probe D3-brane parametrized by a radial coordinate more adequately
can be described by the geometrical tachyon dynamics, as is this
was developed for the open tachyon on an unstable brane.

We find that the dynamics on the D3-brane can be described by a
geometrical tachyon potential modified by a function of the bulk
tachyon. For the particular solution we are considering, this
function alters the height and the shape of the geometrical
tachyon potential. Away from the D3-brane background, the
geometrical tachyon potential asymptotes to a constant value where
$T_{geo}=\infty$ while it goes to zero as the probe brane
approaches the D3-brane background with $T_{geo}=-\infty$,
resembling to the usual geometrical tachyon picture of a decaying
 D3-brane.

There are however important consequences of the presence of the
bulk tachyon field to the cosmological evolution of the
brane-universe. Considering a scalar curvature term on the probe
brane, the cosmological evolution of the brane-universe will be
driven by its own gravitational field and by a ``mirage" matter
coming from the bulk. We show that inflation occurs where the bulk
tachyon condenses at the top of the geometrical tachyon potential.
The condensation of the bulk tachyon occurs at weak string
coupling where the D3-brane bulk solution can be trusted.
Depending on the parameters of the theory, we show that the
brane-universe has also late time acceleration. Also the equation
of state parameter $w$ during the early inflationary phase starts
from the value  -1  and taking the values of the radiation and
matter dominated epochs, it relaxes to -1 at the late accelerating
phase.

The paper is organized as follows. In Sec.~2 we study the movement
of a probe D3-brane in the background of a D3-brane of Type-0
string theory. We parameterize this motion by an open tachyon
field rolling down its potential and we analyze how this motion is
affected by the presence of the bulk tachyon. In Sec.~3 we couple
the geometrical tachyon with four-dimensional gravity on the probe
brane and we follow the cosmological evolution of a brane-universe
driven by its own gravitational field and by the ``mirage" matter
induced on the probe brane by the bulk. Finally, in Sec.~4 we
discuss our results.

\section{A Probe D3-Brane Moving in the Background of D3-Branes with a Bulk Tachyon Field}

The background metric we consider has the general form
\begin{equation}\label{in.met}
ds^{2}_{10}=g_{00}(r)dt^{2}+g(r)(d\vec{x})^{2}+
  g_{rr}(r)dr^{2}+g_{S}(r)d\Omega~.
\end{equation} This metric takes a particular form in the case
 of a stack of Dp-branes which are RR-charged,
and there is also a non-constant dilaton field
\cite{Horowitz:1991cd}
\begin{equation}
ds^{2}_{10}=\frac{1}{\sqrt{H_{p}}}\Big{(}-dt^{2}+(d \vec{x})
^{2}\Big{)} +\sqrt{H_{p}}\,dr^{2}+ \sqrt{H_{p}}\, r^{2}d \Omega
^{2}_{8-p}~, \label{bhmetricdp}
\end{equation}
where $ H_{p}=1+\Big{(} \frac{L}{r}\Big{)} ^{7-p}$~. In this
background the RR field takes the form
\begin{equation}
C_{012...p}=\sqrt{1+\Big{(} \frac{r_{0}}{L}\Big{)} ^{7-p}}\,
\frac{1-H_{p}(r)}{H_{p}(r)}~,
\end{equation} while the dilaton field is given by $ e^{\Phi}=H_{p}^
{(3-p)/4}$. We consider a probe D3-brane moving in this general
background. The dynamics on the probe D3-brane will be described
by the
 Dirac-Born-Infeld plus the Wess-Zumino action,
\begin{eqnarray}\label{B.I. action}
  S_{probe}&=&S_{DBI}+S_{WZ}=\tau~\int~d^{4}\xi \,\,
  e^{-\Phi}\sqrt{-det(\hat{G}_{\alpha\beta}+(2\pi\alpha')F_{\alpha\beta}-
  B_{\alpha\beta})}
  \\  \nonumber
  &+&\tau~\int~d^{4}\xi \,\, \hat{C_{4}}~.
\end{eqnarray}
 The induced metric on the brane is
\begin{equation}\label{ind.metric}
  \hat{G}_{\alpha\beta}=G_{\mu\nu}\frac{\partial x^{\mu}\partial x^{\nu}}
  {\partial\xi^{\alpha}\partial\xi^{\beta}}~,
\end{equation}
 with similar expressions for $F_{\alpha\beta}$ and
 $B_{\alpha\beta}$.
For an observer on the brane the Dirac-Born-Infeld action is the
volume of the brane trajectory modified by the presence of the
anti-symmetric two-form $ B_{\alpha\beta}$, and world-volume
anti-symmetric gauge fields $ F_{\alpha\beta}$.

Assuming that there are no anti-symmetric two-form $
B_{\alpha\beta}$ fields and world-volume anti-symmetric gauge
fields $ F_{\alpha\beta}$, in the static
 gauge, $x^{\alpha}=\xi^{\alpha},\alpha=0,1,2,3 $
 using (\ref{ind.metric}) we can calculate the bosonic part of the
 brane Lagrangian which reads~\cite{Kehagias:1999vr}
\begin{equation}\label{brane Lagr}
\mathcal{L}=\sqrt{A(r)-B(r)\dot{r}^{2}-D(r)h_{ij}\dot{\varphi}^{i}\dot{\varphi}^{j}}
-C(r)~,
\end{equation}
where $h_{ij}d \varphi ^{i} d \varphi^{j}$ is the line
 element of the unit five-sphere, and
 \begin{equation}\label{met.fun}
  A(r)=g^{3}(r)|g_{00}(r)|e^{-2\Phi},
  B(r)=g^{3}(r)g_{rr}(r)e^{-2\Phi},
  D(r)=g^{3}(r)g_{S}(r)e^{-2\Phi},
\end{equation}
where $C(r)$ is the $RR$ background. The problem is effectively
one-dimensional and can be solved easily. The momenta are given by
\begin{eqnarray}
p_{r}&=&-\frac{\dot{r}}{\sqrt{1-H(r)\dot{r}^{2}}}~,\nonumber
\\
p_{i}&=&-\frac{r^{2}h_{ij}\dot{\phi}^{j}}{{\sqrt{1-H(r)\dot{r}^{2}-
H(r)r^{2}h_{ij}\dot{\phi}^{i}\dot{\phi}^{j}}}}~. \nonumber
\end{eqnarray}
Since (\ref{brane Lagr}) is not explicitly time dependent and the
$\phi$-dependence is confined to the kinetic term for
$\dot{\phi}$, for an observer in the bulk, the brane moves in a
geodesic parameterized by a conserved energy $E$ and a conserved
angular momentum $l^{2}$ given by
\begin{eqnarray}
E&=&\frac{\partial\mathcal{L}}{\partial \dot{r}}\dot{r}+
\frac{\partial\mathcal{L}}{\partial
\dot{\phi}^{i}}\dot{\phi}^{i}-\mathcal{L}=p_{r}\dot{r}+p_{i}\dot{\phi}^{i}-\mathcal{L}~,\nonumber
\\ l^{2}&=&h^{ij}\frac{\partial\mathcal{L}}{\partial
\dot{\phi}^{i}} \frac{\partial\mathcal{L}}{\partial
\dot{\phi}^{j}}=h^{ij}p_{i}p_{j}~.\label{equationsCons}
\end{eqnarray}

 In particular, we are interested in the movement of a
probe D3-brane in a specific type of the above background, namely
the Type-0  string background
\cite{Polyakov:1998ju,Klebanov1:1998yy,Klebanov:1998yy} in which,
except from the RR fluxes there is also a bulk tachyon field,
coupled to them.

The action of the Type-0 string is given by
\cite{Klebanov1:1998yy}
\begin{eqnarray}\label{action}
S_{10}&=&~\int~d^{10}x \,\, \sqrt{-g}\Big{[} e^{-2\Phi} \Big{(}
 R+4(\partial_{\mu}\Phi)^{2} -\frac{1}{4}(\partial_{\mu}T_{bulk})^{2}
-\frac{1}{4}m^{2}T^{2}_{bulk}-\frac{1}{12}H_{\mu\nu\rho}H^{\mu\nu\rho}\Big{)}
\nonumber \\&& - \frac{1}{4}f(T_{bulk})|F_{5}|^{2} \Big{]}~,
\end{eqnarray} where $F_{5}=dC_{4}$ is the 5-form field strength
of the RR field. The tachyon is coupled to the RR field via the
function
\begin{equation}\label{ftac}
 f(T_{bulk})=1+T_{bulk}+\frac{1}{2} T^{2}_{bulk}~.
\end{equation}
The bulk tachyon field appearing in (\ref{action}) is a closed
tachyon field which is the result of GSO projection and there is
no spacetime supersymmetry in the theory. The tachyon field
appears in (\ref{action}) via its kinetic term, its potential and
via the tachyon function $f(T_{bulk})$. The potential term is
giving a negative mass squared term which is signaling an
instability in the bulk. However, it was shown in
\cite{Klebanov:1998yy,Klebanov1:1998yy,Klebanov:1999um} that
because of the coupling of the tachyon field to the RR flux, the
negative mass squared term can be shifted to positive values if
the function $f(T_{bulk})$ has an extremum, i.e.
$f^{\prime}(T_{bulk})=0$. This happens in the background where the
tachyon field acquires vacuum expectation value $T_{bulk,~vac}=-1$
\cite{Klebanov:1998yy,Klebanov1:1998yy}. In this background the
dilaton equation is \be \nabla^{2}\Phi=-\frac{1}{4\alpha^{\prime}}
e^{\frac{1}{2}\Phi}T^{2}_{bulk,~vac}~. \ee This equation is giving
a running of the dilaton field which means that the conformal
invariance is lost, and $AdS_{5}\times S^{5}$ is not a solution.
Therefore, the closed tachyon condensation is responsible for
breaking the 4-D conformal invariance of the theory.  However, the
conformal invariance is restored in two conformal points,
corresponding to IR and UV fixed points, when the tachyon field
gets a constant value. The flow from IR to UV as exact solutions
of the equations of motion derived from action (\ref{action}) is
not known, only approximate solutions exist
\cite{Klebanov:1998yy,Minahan:1998tm,Grena:2000xw}. In these
solutions, the closed tachyon field starts in the UV from
$T_{bulk}=-1$, grows to larger values and passing from an
oscillating phase it reaches $T_{bulk}=0$ at the IR. However, if
the dilaton and tachyon fields are constant, an exact solution of
a D3-brane can be found \cite{Klebanov1:1998yy,Klebanov:1998yy}
 \begin{equation}
ds^{2}_{10}=\frac{1}{\sqrt{H}}\Big{(}-dt^{2}+(d \vec{x})
^{2}\Big{)} +\sqrt{H}\,dr^{2}+ \sqrt{H}\, r^{2}d \Omega ^{2}_{5}~,
\label{bhmetrictyp0}
\end{equation}
where $ H=1+\Big{(} \frac{e^{\Phi_{0}}Q}{2r^{4}}\Big{)}$, Q is the
electric RR charge and $ \Phi_{0}$ denotes a constant value of the
dilaton field. If we define
$L=\Big{(}e^{\Phi_{0}}Q/2\Big{)}^{1/4}$, $H$ can be rewritten as $
H=1+\Big{(} \frac{L}{r}\Big{)}^{4}$. In the Type-0 background we
have
\begin{equation}\label{met.fun1}
  A(r)=e^{-\Phi_{0}}k(T_{bulk})^{2}H^{-2}(r)~,
  B(r)=e^{-\Phi_{0}}k(T_{bulk})^{2}H^{-1}(r)~,
  D(r)=e^{-\Phi_{0}}k(T_{bulk})^{2}H^{-1}(r)r^2~,
\end{equation}
where the function $k(T_{bulk})$ appears because of the presence
of the tachyon field in the bulk. Its form it is found to be
\cite{Klebanov:1998yy,Garousi:1998fg,Garousi:1999fu}
\begin{equation}
k(T_{bulk})=1+\frac{T_{bulk}}{4}+\frac{3T_{bulk}^{2}}{32}+....\label{kfunct}
\end{equation}

 Then solving equations (\ref{equationsCons})  for
$\dot{r}$ and $ \dot{\phi}$ we find \bea
\dot{r}^{2}&=&H^{-1}-\frac{H^{-2}}{(C+E)^2}(\alpha^{2}H^{-1}+\sqrt{H-1}\frac{l^2}{L^2})~,
\,\,~~ \label{functionr1}\\
h_{ij}\dot{\varphi}^{i}\dot{\varphi}^{j}&=&\frac{H^{-2}(H-1)}{(C+E)^{2}}\frac{\ell^{2}}{L^4}~.
\eea Putting these expressions back to (\ref{brane Lagr}) the DBI
action becomes
 \begin{equation}
 S_{probe}=\tau\int d^4x \,\Big{[}\, k(T_{bulk})H^{-1}\sqrt{1-\Big{(} 1-\frac{\alpha^{2}}{(C+E)^{2}}H^{-2}
 \Big{)}}-\frac{1}{\tau}C(r)\Big{]}\label{dbi1}~,
  \end{equation}
where $\alpha=e^{-\Phi_{0}}k(T_{bulk})$ and $C(r)=\tau
f^{-1}(T_{bulk})H^{-1}+Q_{1}$ with $Q_{1}$ an integration constant
\cite{Papantonopoulos:2000yz} and the factor $e^{-\Phi_{0}}$ has
been absorbed in the coupling $\tau$.  Note that there is no
world-volume coupling of the bulk tachyon field to the RR fields
\cite{Garousi:1999fu}. Absorbing the integration constant $Q_{1}$
in the energy $E$, the action (\ref{dbi1}) becomes
\begin{equation}
S_{probe}=\tau\int d^4x \,\,
k(T_{bulk})H^{-1}\Big{[}\sqrt{1-\Big{(}
1-\frac{\alpha^{2}}{(C+E)^{2}}H^{-2}
 \Big{)}}-f^{-1}(T_{bulk})k^{-1}(T_{bulk})\Big{]}\label{dbi2}~.
 \end{equation}

 The motion of the probe brane in a geodesic with a
 conserved energy $E$ and a conserved angular momentum $l^{2}$,
  can be parameterized by a single scalar field $T_{geo}$ if
 we define
 \begin{equation}
 T_{geo}=\int dH \Big{(}-\frac{L}{4}H^{1/2}(H-1)^{-5/4}\Big{)}
 \sqrt{1-\frac{l^2}{L^2} \frac{H(H-1)^{1/2}}{\alpha^{2}}
 \Big{[}1+\frac{l^2}{L^2} \frac{H(H-1)^{1/2}}{\alpha^{2}} - \frac{H^{2}(C+E)^{2}}{\alpha^{2}} \Big{]}^{-1}}
 \label{fieldredf}~.
 \end{equation}
 This relation gives an explicit relation between the radial
 coordinate $r$ and the field $T_{geo}$. In the case of $l=0$, $T_{geo}$ can be expressed as
 a hypergeometric function of $H(r)$ \cite{Panigrahi:2004qr,Saremi:2004yd}
\begin{equation}
T_{geo}(H)= \frac{LH^{1/2}}{(H-1)^{1/4}}\Big{(} 1-(H-1)^{1/4}
~_{2}F_{1}(\frac{1}{2},\frac{1}{4};\frac{3}{2}; H)
\Big{)}~.\label{hyper}
\end{equation}
  Therefore, the scalar
 field $T_{geo}$ has a clear geometrical interpretation in term of
 the coordinate $r$, the distance of the probe brane from the
 D3-branes in the bulk. Therefore, the movement of the probe D3-brane can be parametrized in the same way
 as in the geometrical tachyon picture~\cite{Kutasov:2003er,Kutasov:2004dj,Kutasov:2004ct}
  with a open tachyon field rolling
 down its potential. To see this, using  the field redefinition
 (\ref{fieldredf}) the action (\ref{dbi2}) becomes
 \begin{equation} \label{TachyonAction}
 S_{probe}= \int d^{4}x \,\, V(T_{geo}) \Big{[}\sqrt{1+g_{\alpha \beta}\partial^{\alpha}T_{geo}
 \partial^{\beta}T_{geo}} -f^{-1}(T_{bulk})k^{-1}(T_{bulk}) \Big{]}\label{dbi3}~,
 \end{equation}
where the only nontrivial component of the metric $g_{\alpha
\beta}$ is the time component and the open tachyon potential is
given by
 \begin{equation}
V(T_{geo})=\frac{\tau k(T_{bulk})}{H(T_{geo})}~.\label{potent}
\end{equation}
To get the explicit form of $V(T_{geo})$ we must invert the
expression (\ref{fieldredf}) to get $H(T_{geo})$. The term \be
\int d^{4}x \,\, V(T_{geo})f^{-1}(T_{bulk})k^{-1}(T_{bulk})\ee in
(\ref{TachyonAction}), is the  Wess-Zumino term which is a
function of the geometrical tachyon field and the projection of
the RR field of the bulk on the brane. We can compare it with the
usual form of the Wess-Zumino term of an unstable D-brane in the
background of RR
fields~\cite{Billo:1999tv,Kennedy:1999nn,Sen:2003tm,Okuyama:2003wm,Kluson:2005fj}
\begin{equation}
S_{WZ}=\int W(T) dT \wedge \mathbf{C}\wedge e^{\mathbf{F}}~,
\end{equation} where $\mathbf{F}$ is a gauge field on the brane,
$W(T)$ a function of open tachyon field and $\mathbf{C}$ a
combination of the RR fields of the bulk. In our case we do not
have an explicit gauge field on the probe brane, the functions
$f^{-1},k^{-1}$ of the bulk tachyon field correspond to the field
$\mathbf{C}$, while the potential (\ref{potent}) corresponds to
the term $W(T) dT$. The derivative of the open tachyon field
arises because the function $H(T_{geo})$ in (\ref{potent}) can be
express as $dT_{geo}$ (see (\ref{fieldredf}) and (\ref{diffeq1}),
(\ref{diffeq2})).

The form of the tachyonic action (\ref{TachyonAction}) indicates
that the movement of the probe D3-brane in the field of other
D3-branes, can be described by an open tachyon field rolling down
its potential
\cite{Sen:2002nu,Sen:2002in,Sen:2002an,Sen:2002vv,Sen:2002qa}.
However, the open tachyon field in our case is a "mirage`` tachyon
field and it is not a direct consequence of a D3-brane
instability. The novel feature here is the presence of the bulk
tachyon field $T_{bulk}$. The bulk tachyon field appears in the
potential (\ref{potent}) and if $l\neq 0$ it also appears in the
definition of $T_{geo}$ in (\ref{fieldredf}). In the next two
subsections we will study what is its effect on the movement of
the D3-brane.

\subsection{A Probe D3-Brane Moving Radially in the Background of
Other D3-Branes}

We will consider the probe D3-brane moving radially in this
non-conformal string background. The tachyon field $T_{geo}$ is
related to the distance $r$ of the probe brane from the bulk
D3-branes via the following equation \be \frac{dT_{geo}}{dr} =
\sqrt{H(r)} = \sqrt{1+\frac{L^4}{r^4}} \label{diffeq1}~, \ee which
is easily derived from (\ref{fieldredf}) (for $l=0$) and its
potential is given by \be V(T_{geo}) = \frac{\tau
k(T_{bulk})}{H(r)}~. \label{pot1} \ee Because of (\ref{diffeq1}),
the relation between the radial mode and $T_{geo}$ is not
logarithmic like in the case of NS5 background
\cite{Kutasov:2004dj} or D5-brane background
\cite{Burgess:2003mm,Panigrahi:2004qr,Saremi:2004yd,Kluson:2005qx,Kluson:2005jr}
but rather polynomial, which means that the explicit form of
$T_{geo}(r)$ can not be found. Indeed, $T_{geo}(r)$ is a
combination of elliptic integrals of the first and the second
kind. Nevertheless asymptotically the differential equation
(\ref{diffeq1}) can be evaluated. In the infrared limit $(r
\rightarrow 0)$ we find that \bea T_{geo}(r)
&\sim& - \frac{L^2}{r} \rightarrow - \infty \label{T1}\\
\frac{1}{\tau}V(T_{geo}) &\sim& k(T_{bulk})\frac{L^4}{T_{geo}^4}
\rightarrow 0~, \label{V1} \eea whereas in the ultra violet limit
$(r \rightarrow \infty)$ \bea
T_{geo}(r) &\sim& r \rightarrow \infty \label{T2}\\
\frac{1}{\tau}V(T_{geo}) &\sim&
k(T_{bulk})\Big{(}1-\frac{L^4}{T_{geo}^4} \Big{)} \simeq const.
\label{V2} \eea In the regime $T_{geo} \rightarrow \infty$ we see
that the interaction potential (\ref{V2}) goes like $-1/T_{geo}^4
\simeq -1/r^4$. Consequently we do not recover the standard long
range gravitational attraction between the probe D3-brane and the
background D3-branes but a modified long range attraction. In the
opposite regime $(T_{geo} \rightarrow -\infty)$, during the
"Radion Matter" phase when the probe D3-brane is close to the
background D3-branes, we find that the potential $V(T_{geo})$ in
(\ref{V1}) goes to zero as $1/T_{geo}^4$. The transition between
the two behaviours occurs at $r \sim L$. Note that the limits
(\ref{T1}) and (\ref{T2}) can also be obtained from the asymptotic
behaviour of the hypergeometric function (\ref{hyper}).

The presence of the bulk tachyon field through the function
$k(T_{bulk})$ in (\ref{pot1}) influences the shape of the
geometrical tachyon potential. The height of the open tachyon
potential in its maximum value in a non-BPS brane is equal to the
tension of the D3-brane. In our case, in the UV fixed point where
the bulk tachyon field condenses, $k(T_{bulk})=3/4$, indicating
that the presence of the bulk tachyon is lowering the D3-brane
tension. The bulk tachyon field changes from -1 in the UV to 0 in
the IR, therefore the geometrical tachyon potential~(\ref{pot1})
alters its shape as the geometrical tachyon rolls down to its
minimum. As we will see in the next section this has important
consequences in the cosmological evolution of the brane-universe.

In the open tachyon case of an unstable D3-brane the condensation
of the open tachyon field exactly cancels the probe D3-brane
tension and then we do not have any perturbative open string
states in the spectrum. Then, the exchange of massless closed
strings dominate. The same happens here because $k(T_{bulk})=1$ in
the infrared and because $V(T_{geo})=0$  they are not open string
states. However in our case because they are closed string states
in the spectrum the low-energy approximation may breaks down. This
is certainly the case when one considers the full system of
equations resulting from the action (\ref{action}). There are
approximate solutions of these equations with non-constant tachyon
and non-constant dilaton fields
\cite{Klebanov:1998yy,Minahan:1998tm}, where the dilaton field in
the infrared gets large values and as a consequence, the string
effective coupling becomes large. The solutions
(\ref{bhmetrictyp0}) and we are considering in our analysis, have
a constant dilaton field and the string coupling can be choosen to
be small so the low-energy approximation can be trusted.

 Demanding  $\dot{r} \geq 0$ and using (\ref{functionr1}) for $l=0$, we get the
 inequality
  \be H^2+2 \frac{\tau e^{\Phi_{0}} \beta}{E^2} H
\geq \frac{\tau e^{\Phi_{0}}(\alpha^2(T_{bulk}) -
\beta^2(T_{bulk}))}{E^2}~, \label{constraint1} \ee where
$\beta(T_{bulk}) = e^{-\Phi_{0}}f^{-1}(T_{bulk})$. In the limit
where $r \ll L$, this condition becomes \be \frac{L^8}{r^8}
\Big{(}1+2\frac{\tau e^{\Phi_{0}} \beta}{E^2} \frac{r^4}{L^4}
\Big{)} \geq \frac{\tau e^{\Phi_{0}}(\alpha^2(T_{bulk}) -
\beta^2(T_{bulk}))}{E^2} \label{constraint2}\ee  and for $-1 \leq
T_{bulk} \leq 0$ \be \alpha^2(T_{bulk})-\beta^2(T_{bulk}) =
e^{-2\Phi_{0}} \Big{(}k^2(T_{bulk}) - f^{-2}(T_{bulk}) \Big{)}
\leq 0~. \ee Therefore the constraint (\ref{constraint2}) is
always satisfied. Consequently the probe D3-brane is free to
approach the background D3-branes at any distance. In the limit
where $r \gg L$, the condition (\ref{constraint1}) becomes \be 1+2
\frac{\tau e^{\Phi_{0}}\beta}{E^2} \geq \frac{\tau
e^{\Phi_{0}}(\alpha^2(T_{bulk}) - \beta^2(T_{bulk}))}{E^2}~,
\label{constraint3}\ee which is also always satisfied. As a
consequence the probe D3-brane can escape to infinity.

\subsection{ A Spinning Probe D3-Brane in the Background of
D3-Branes}

In the case of a non-zero angular momentum, the tachyon field
$T_{geo}$ is related to the distance $r$ of the probe brane from
the background D3-branes via the following  equation \be
\frac{dT_{geo}}{dr} = H^{1/2}
 \sqrt{1-\frac{l^2}{L^2} \frac{H(H-1)^{1/2}}{\alpha^{2}}
 \Big{[}1+\frac{l^2}{L^2} \frac{H(H-1)^{1/2}}{\alpha^{2}} - \frac{H^{2}(C+E)^{2}}{\alpha^{2}} \Big{]}^{-1}}
 \label{diffeq2}. \ee The explicit
form of $T_{geo}(r)$ can not be found. As previously, we can
asymptotically find the solutions of the differential equation
(\ref{diffeq2}) modified by the angular momentum. In the infrared
limit $(r \rightarrow 0)$ we find that \be T_{geo}(r) \sim -
\frac{L^2}{r} + \frac{l^2}{L^2} \frac{r}{2E} \rightarrow - \infty
\label{T3}~, \ee which gives \bea r(T_{geo}) &\sim&
-\frac{L^2}{T_{geo}}+\frac{1}{2}
 \frac{l^2 L^2}{E \,\, T_{geo}^4}\\
\frac{1}{\tau}V(T_{geo}) &\sim& -k(T_{bulk})\frac{L^4}{T_{geo}^4}
\Big{(}1-\frac{2l^2}{E \,\,T_{geo}^3} \Big{)} \rightarrow 0~,
\label{V3} \eea whereas in the ultra violet limit $(r \rightarrow
\infty)$ \be T_{geo}(r) \sim r + \frac{l^2}{2 \alpha^4}
\frac{1}{r}
\rightarrow \infty \label{T4}\ee and we obtain \bea r(T_{geo}) &\sim& T_{geo} -\frac{l^2}{2 \alpha^4T_{geo}}\\
\frac{1}{\tau}V(T_{geo}) &\sim&
k(T_{bulk})\Big{(}1-\frac{L^4}{T_{geo}^4}-\frac{2l^2 \,\,
L^4}{\alpha^4 \,\, T_{geo}^6} \Big{)} \simeq const. \label{V4}
\eea Therefore we have the same behaviour of the geometrical
tachyon field rolling down its potential as in the previous case
of the radial motion.

 The constraint $\dot{r} \geq
0$ using (\ref{functionr1}) can be rewritten as \be H^2+2
\frac{\tau e^{\Phi_{0}}\beta}{E^2} H - \frac{\tau
e^{\Phi_{0}}l^2}{E^2 r^2} H \geq \frac{\tau
e^{\Phi_{0}}(\alpha^2(T_{bulk}) - \beta^2(T_{bulk}))}{E^2}~.
\label{constraintl1} \ee In the limit where $r \ll L$ this
condition becomes \be \frac{L^8}{r^8} \Big{(}1-\frac{\tau
e^{\Phi_{0}}l^2}{E^2 L^2} \frac{r^2}{L^2}+2\frac{\tau
e^{\Phi_{0}}\beta}{E^2} \frac{r^4}{L^4} \Big{)} \geq \frac{\tau
e^{\Phi_{0}}(\alpha^2(T_{bulk}) - \beta^2(T_{bulk}))}{E^2}~.
\label{constraintl2}\ee while when $r \gg L$ we get \be
1-\frac{\tau e^{\Phi_{0}}l^2}{E^2 L^2} \frac{L^2}{r^2}+2
\frac{\tau e^{\Phi_{0}}\beta}{E^2} \geq \frac{\tau
e^{\Phi_{0}}(\alpha^2(T_{bulk}) - \beta^2(T_{bulk}))}{E^2}~.
\label{constraintl3}\ee In both cases the constraint
$\dot{r}^2\geq0$ is always satisfied. The probe D3-brane is again
free to move in the whole background.

\section{Coupling to Gravity-Tachyon Cosmology}

The rolling tachyon describing the time evolution of a decaying
D-brane in open string theory, initiated the development of the
tachyon cosmology with the hope that the open tachyon field on an
unstable brane plays the r$\hat{o}$le of the scalar field driving
the inflation \cite{tachyon-cosmology}. However, it seems unlikely
that the tachyon field is responsible for inflation. The reason is
that this scenario is plagued by serious problems like
incompatibility of slow-roll, too steep potential and the lack of
a mechanism for reheating \cite{Kofman:2002rh,tachcosmo}.

In a parallel development, the study of the dynamics of the
rolling tachyon field was extended to cosmological systems in the
case of a D3-brane propagating in the background of other
coincident NS5 branes \cite {Yavartanoo:2004wb}. A formulation of
tachyon inflation was proposed in \cite{Thomas:2005fu} using the
geometrical tachyon arising from the time dependent motion of a
D3-brane in the background geometry due to parallel NS5-branes
arranged around a ring. A phenomenological analysis was performed
on this model, constraining the various parameters from the recent
observational data \cite{Panda:2005sg}. The further study of
various configurations and backgrounds may give us more
information on the relevance of the geometrical tachyon to
cosmology.

In this section we will study the influence of a bulk tachyon
field to the cosmological evolution a probe D3-brane as it moves
in the gravitational field of other D3-branes of Type-0 using the
the geometrical tachyon picture. On the probe D3-brane we will
introduce a four-dimensional scalar curvature
term~\cite{Dvali:2000hr}.

To capture the dynamics of the induced gravitational field on the
brane, assuming that it is minimally coupled, we consider the
  DBI action of the geometrical tachyon field coupled to gravity
  \be \label{CosmoTachyonAction} S_{cosmo} =
\int{d^{4}x \sqrt{-g} \Big{(} \frac{R}{16 \pi G} - V(T_{geo})
\sqrt{1+g^{\mu \nu}
\partial_{\mu}T_{geo} \partial_{\nu}T_{geo}}~ \Big{)}}~. \ee
In the above action the geometrical tachyon field acts as a local
matter source on the brane and its energy-momentum tensor is given
by
 \be T_{\mu \nu}=
\frac{-2}{\sqrt{-g}} \frac{\partial{S_{cosmo}}}{\partial{g^{\mu
\nu}}} = V(T_{geo}) \frac{\partial_{\mu}T_{geo}
\partial_{\nu}T_{geo}}{\sqrt{1-\dot{T}_{geo}^2}} - V(T_{geo})g_{\mu \nu}
\sqrt{1-\dot{T}_{geo}^2}~. \ee Assuming that the tachyon field is
described by a homogeneous fluid with $T_{\mu \nu} = pg_{\mu \nu}
+ (p+\rho)u_{\mu}u_{\nu}$ we can define the following energy
density and pressure \bea
\label{Rho} \rho_{tch} &=& \frac{V(T_{geo})}{\sqrt{1-\dot{T}_{geo}^2}}~, \\
\label{pressure} p_{tch} &=& -V(T_{geo})
\sqrt{1-\dot{T}_{geo}^2}~. \eea For the string background we are
considering, from equations (\ref{dbi2}) and (\ref{TachyonAction})
we have that \be
\sqrt{1-\dot{T}_{geo}^2}=\frac{\alpha(T_{bulk})}{H}\Big{(}E+\frac{\beta(T_{bulk})}{H}\Big{)}^{-1},
\label{brel} \ee and using (\ref{pot1}) with (\ref{Rho}) and
(\ref{pressure}) we obtain the following expressions for the
energy density and the pressure \bea
\frac{\rho_{tch}(H)}{\tau} &=& E + \frac{\beta(T_{bulk})}{H}~,\label{rtch}\\
\frac{p_{tch}(H)}{\tau} &=&
-\frac{\alpha^2(T_{bulk})}{H^2}\Big{(}E+\frac{\beta(T_{bulk})}{H}\Big{)}^{-1}\label{pres}~.
\eea One should keep in mind here, that in our approach the
geometrical tachyon field we are considering, because of the
identification we did in (\ref{fieldredf}) is a mirage tachyon
field on the probe brane, in the sense that it is an expression of
bulk quantities. For this reason, we use relation (\ref{brel})
which is not derived from the action (\ref{CosmoTachyonAction}).
Also observe that as the brane moves towards the bulk D3-branes,
$\rho_{tch}$ and $p_{tch}$ of (\ref{Rho}) and (\ref{pressure}) are
changing, being functions of $r$. The presence of the scalar
curvature term in the action (\ref{CosmoTachyonAction}), assuming
a flat Robertson-Walker metric on the brane, leads to the
Friedmann equation \be \mathcal{H}^2_{tch}=\frac{8\pi
G}{3}\rho_{tch}~, \label{friedtch}\ee which gives a cosmological
evolution  because of the presence of the gravitational field on
the brane.

As the probe brane is moving in the background string theory, it
will also experience the effect of the bulk gravitational field.
This effect can be calculated with the techniques of mirage
cosmology~\cite{Kehagias:1999vr,Papantonopoulos:2000yz,mirage-cosmology}.
The presence of the Type-0 string background induces on the probe
brane~\cite{Papantonopoulos:2000yz}, a four-dimensional metric
 \begin{equation}
d\hat{s}^{2}=-\frac{H^{-5/2}\alpha(T_{bulk})^{2}}{\Big{(}\beta(T_{bulk})H^{-1}+E\Big{)}^{2}}dt^{2}+g(d\vec{x})^{2}.
\end{equation} Defining  the cosmic time $\eta$  as
\begin{equation}\label{cosmic}
 d\eta=\frac{H^{-5/4}\alpha(T_{bulk})}{\Big{(}\beta(T_{bulk})H^{-1}+E\Big{)}}dt~,
\end{equation} the induced metric becomes
\begin{equation}\label{fin.ind.metric}
d\hat{s}^{2}=-d\eta^{2}+g(r(\eta))(d\vec{x})^{2}.
\end{equation}
 The induced metric on the brane (\ref{fin.ind.metric}) is the standard form of a flat
expanding universe.  Defining $g=a$ we
get~\cite{Papantonopoulos:2000yz}
\begin{equation}\label{dens}  \Big{(}\frac
{\dot{a}}{a}\Big{)}^{2}= \frac{\tau}{L^{2}}\Big{(}
\frac{H-1}{H}\Big{)}^{5/2}\Big{(}\alpha(T_{bulk})^{-2}(\beta(T_{bulk})+EH)^{2}-1
\Big{)}~,
\end{equation}
 where the dot stands for derivative with respect to cosmic time. The
 right hand side of (\ref{dens}) can be interpreted in terms of an
 effective or ``mirage" matter density on the probe brane
 \begin{equation}\label{rmir} \rho_{mir}=
\frac{3\tau}{8\pi G L^{2}}\Big{(}
\frac{H-1}{H}\Big{)}^{5/2}\Big{(}\alpha(T_{bulk})^{-2}H^{2}\,\Big{(}E+\frac{\beta(T_{bulk})}{H}\Big{)}^{2}-1
\Big{)}~.
\end{equation} We can also calculate the ``mirage" pressure using  \begin{eqnarray}  \label{dadot}
\frac{\ddot{a}}{a}&=& \Big{[}1+\frac{1}{2}a\frac{\partial}
{\partial a}\Big{]}\frac{8\pi G}{3}\rho_{mir}
\end{eqnarray}
and setting  the above equal to $-\frac{4\pi G
}{3}(\rho_{mir}+3p_{mir})$, we can define the ``mirage" pressure
 \bea p_{mir}&=&-\frac{\tau}{\kappa\,L^2}\Big{(}\frac{H-1}{H}\Big{)}^{5/2}
\Big{(}\Big{(}\frac{3H-13}{H-1}\Big{)}\Big{(}\alpha(T_{bulk})^{-2}
H^2\Big{(}E+\frac{\beta(T_{bulk})}{H}\Big{)}^{2}-1\Big{)}\nonumber \\
&-&8E\,\alpha(T_{bulk})^{-2}H^2\Big{(}E+\frac{\beta(T_{bulk})}{H}\Big{)}\Big{)}~.
\label{pmir}\eea Then, as the brane is moving in the gravitational
field of the bulk, because of this motion~~\cite{Kehagias:1999vr},
there will be a cosmological evolution on the brane described by
the Friedmann equation \be \mathcal{H}^2_{mir}=\frac{8\pi
G}{3}\rho_{mir}~. \label{friedmir}\ee

\begin{figure}
\begin{center}
\includegraphics[scale=1.5]{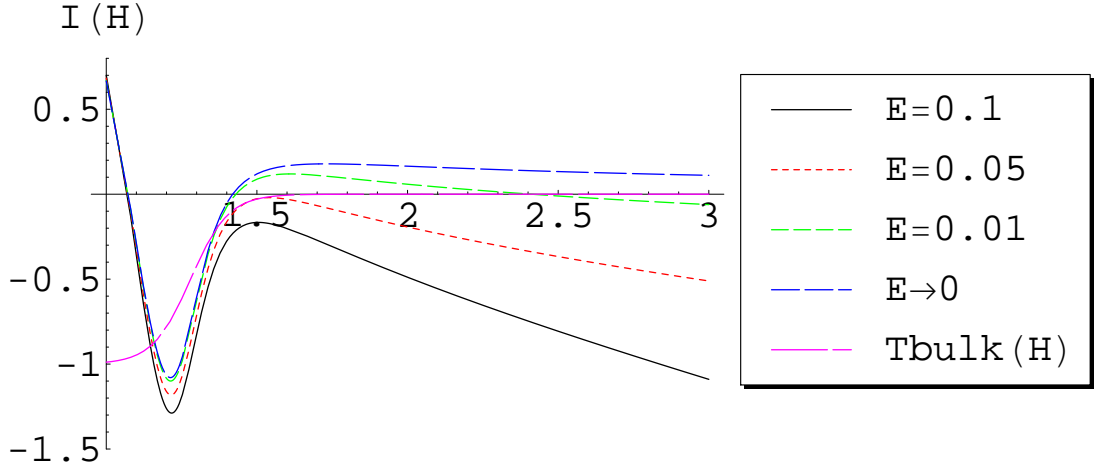}
\caption{The inflation parameter $I(H)$ as a function of $H$ for
$\zeta=8$, $\sigma=1.28$ and for different values of $E$ in $\tau$
units.} \label{InflationPlot}
\end{center}
\end{figure}

Therefore, as the geometrical tachyon rolls down its potential it
feels two effects: the gravitational field of its own and the
gravitational field of the bulk D3-branes. Our basic assumption of
the probe approximation allows us to assume, because the D3-brane
as it moves does not backreact with the background, that the above
two contributions give an additive effect on the cosmological
evolution of the probe brane,  and hence it is described by the
Friedmann equation \be \mathcal{H}^2_{probe}= \frac{8\pi
G}{3}\Big{(}\rho_{tch}+\rho_{mir} \Big{)}\label{fried}~. \ee Then,
using (\ref{rtch}) and (\ref{rmir}), the Friedmann equation
(\ref{fried}) becomes \be \mathcal{H}^2_{probe}=\frac{8\pi G
\tau}{3}\Big{[}\Big{(}E+\frac{\beta(T_{bulk})}{H}\Big{)}
+\frac{3}{8\pi G\,L^2}\Big{(}\frac{H-1}{H}\Big{)}^{5/2}
\Big{(}\alpha(T_{bulk})^{-2}H^{2}\,\Big{(}E+\frac{\beta(T_{bulk})}{H}\Big{)}^{2}-1\Big{)}\Big{]}~.\label{1fried}\\
\ee Also the Raychaudhury equation can be calculated to be \bea
\dot{\mathcal{H}}_{Total}&=&-4\pi G
\tau\frac{(H-1)^{5/4}}{H^{5/2}}
\Big{\{}\frac{4\sqrt{3}}{3\alpha(T_{bulk})}\frac{\beta(T_{bulk})H^{5/4}}{\sqrt{8\pi
G\,L^2}}\,\Big{(}E+\frac{\beta(T_{bulk})}{H}\Big{)}^{1/2}\nn
\\ &~&\times\Big{[}1-\frac{\alpha(T_{bulk})^{2}}
{H^{2}}\Big{(}E+\frac{\beta(T_{bulk})}{H}\Big{)}^{-2}\Big{]}^{1/2}
\nonumber \\
&+&\frac{5}{4\pi
G\,L^2}\Big{(}H-1\Big{)}^{1/4}\Big{[}\frac{H^{2}}{\alpha(T_{bulk})^{2}}\Big{(}E+\frac{\beta(T_{bulk})}
{H}\Big{)}^{2}-1\nn
\\
&+&\frac{4}{5}E\,\frac{H^2}{\alpha(T_{bulk})^2}\Big{(}H-1\Big{)}\Big{(}E+\frac{\beta(T_{bulk})}{H}\Big{)}\Big{]}\Big{\}}~,
\label{2fried}\eea where the time derivative is in the cosmic
time. The inflationary parameter $I(H)$ using (\ref{1fried}) and
(\ref{2fried}) can be defined as \bea I(H)&=&\mathcal{H}^2_{probe}
+\dot{\mathcal{H}}_{probe}^{2}=\nn \\
 &-&\frac{4\pi G \tau}{3}\frac{(H-1)^{5/4}}{H^{5/2}}
\Big{\{}\frac{4\sqrt{3}}{\alpha(T_{bulk})}\,\frac{\beta(T_{bulk})H^{5/4}}{\sqrt{8\pi
G \,L^2}}\,\Big{(}E+\frac{\beta(T_{bulk})}{H}\Big{)}^{1/2}\nn \\
&~&\times\Big{[}1-\frac{\alpha(T_{bulk})^{2}}{H^{2}}\Big{(}E+\frac{\beta(T_{bulk})}{H}\Big{)}^{-2}\Big{]}^{1/2}
\nonumber \\
&-&2
\Big{(}E+\frac{\beta(T_{bulk})}{H}\Big{)}\frac{H^{5/2}}{(H-1)^{5/4}}\Big{[}1-\frac{3}
{2\pi G\,L^2}\,\,\frac{E\,(H-1)^{5/2}}{\alpha(T_{bulk})^{2}\,H^{1/2}}\Big{]} \nonumber \\
&-&\frac{3}{4\pi
G\,L^2}(H-1)^{1/4}\Big{[}\frac{H^{2}}{\alpha(T_{bulk})^{2}}
\Big{(}E+\frac{\beta(T_{bulk})}{H}\Big{)}^{2}-1\Big{]}\Big{(}H-6\Big{)}\Big{\}}~.
\eea The inflationary parameter $I(H)$ depends on the value of
$T_{bulk}$. As we discussed before, we do not know the exact
variation of the bulk tachyon field from UV to IR. The existing
approximate solutions are valid only in the vicinity of the fixed
points and they can not give us a cosmological evolution from
large to small distances. However, we know that the bulk tachyon
field varies from the UV value $T_{bulk}=-1$ to the IR value
$T_{bulk}=0$. We will simulate then this variation with a smooth
function \be
T_{bulk}(H)=\frac{1}{2}\Big{[}\textnormal{Tanh}\Big{(}{\zeta(H-\sigma)}\Big{)}-1\Big{]}~,
\ee where the parameter $\zeta$ controls how steep is the
variation, while $\sigma$ indicates when the transition from -1 to
0 occurs. Using this function, in Fig.~\ref{InflationPlot} we have
plotted the inflationary parameter as a function of $H$ and for
various values of the energy $E$ (in units of $\tau$). The
cosmological evolution of the brane-universe starts with an early
inflationary phase near the value $T_{bulk}=-1$, where the bulk
tachyon field condenses and the string coupling is weak, and as
the bulk tachyon field gets larger values, there is an exit from
inflation and a late acceleration phase as the bulk tachyon field
approaches $T_{bulk}\rightarrow 0$.

We can see the cosmological evolution on the brane-universe using
the geometrical tachyon picture. As we discussed in the Sect. 2-1,
the geometrical tachyon rolling down its potential has two
different asymptotic behaviours. At the UV it starts with
$T_{geo}=\infty$ at the top of the potential, and rolls down to
$V(T_{geo})=0$ in the IR where $T_{geo}=-\infty$. The transition
to the two regimes occurs where $r=L$ or $H=2$. On the other hand,
the background string solution (\ref{bhmetrictyp0}), is reliable
near the UV and IR fixed points in which $H=1$ and
$H\rightarrow\infty$ respectively. As we can see in
Fig.~\ref{InflationPlot}, there is a choice of parameters for
which the early inflationary phase and the exit from it occurs
around $H=1$ which corresponds to the top of the geometrical
tachyon potential. The late inflationary phase occurs for large
$H$ values, where the bulk tachyon field has decoupled, and this
happens in the bottom of the geometrical tachyon potential.

\begin{figure}
\begin{center}
\includegraphics[scale=1.5]{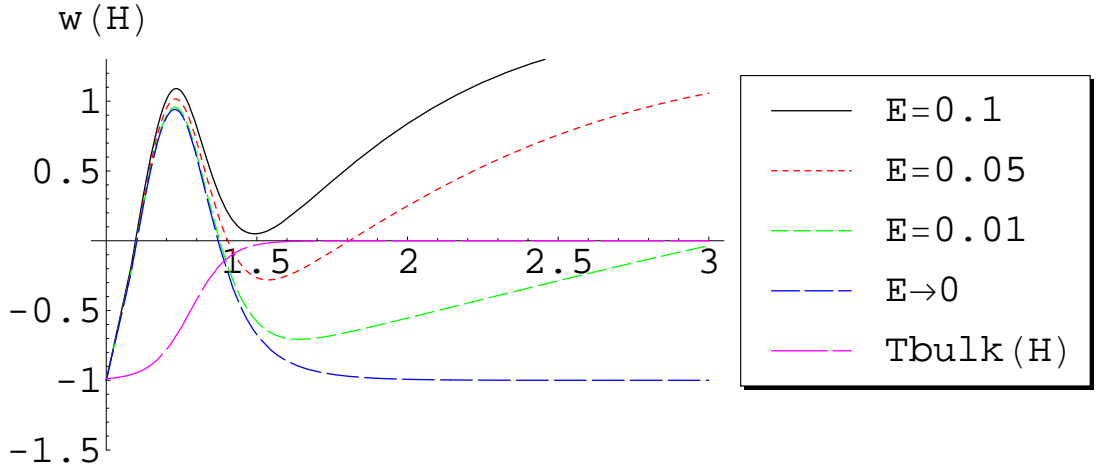}
\caption{Equation of state parameter $w(H)$ as a function of $H$
for $\zeta=8$, $\sigma=1.28$ and for different values of $E$ in
$\tau$ units.} \label{wPlot}
\end{center}
\end{figure}

We can also define the equation of state parameter $w(H)=p_{probe
}/\rho_{probe}$. Then, using (\ref{rtch}), (\ref{pres}),
(\ref{rmir}) and (\ref{pmir}) we can plot $w$ as a function of $H$
and for various values of the energy. We see in Fig.~\ref{wPlot}
that it starts with the value $w=-1$ in the early inflationary
phase, then it gets positive values and finally relaxes again to
$w=-1$ in the late accelerating phase. This picture is appealing,
because the whole cosmological evolution is driven by the dynamics
of the theory, without any dark matter or dark energy.

If we switch off the gravitational field on the moving brane then,
the mirage effect~\cite{Papantonopoulos:2000yz}  gives only the
late accelerating  phase as the probe brane is approaching the
bulk. If we switch off the mirage contribution then the
brane-universe has the early inflationary phase, where the tachyon
field condenses, near the top of the geometrical tachyon
potential. If we decouple the bulk tachyon field from the start,
having a probe brane moving in the background of other D3-branes
then, we have a very short inflationary period in the top of the
geometrical tachyon potential. This can be attributed to the
presence of the gravitational field on the brane.

\section{Conclusions and Discussion}

We have studied the movement of a probe D3-brane in the background
of a D3-brane solution of Type-0 string theory. We have shown that
this movement can be described by the rolling of the geometrical
tachyon down its potential. We found that the presence of the bulk
tachyon of the Type-0 theory, modifies the potential of the
geometrical tachyon. This modification changes the height and the
shape of the potential. Solving the classical equations of motion,
we analysed the influence of the bulk tachyon to the geometrical
tachyon dynamics as it rolls down its potential.

We also studied the cosmological evolution of the brane-universe
on the D3-brane as it moves in this background string theory. We
introduced a scalar curvature term on the moving brane and we
showed that the geometrical tachyon field rolls down its potential
under the influence of its own gravitational field and the
gravitational field of the background. This results in a
cosmological evolution on the moving brane.

We found that there is a choice of parameters, for which early
time inflation occurs where the bulk tachyon field condenses at
weak coupling. This coincides with the UV fixed point of the RG
flows of the bulk Type-0 theory. At this limit, the string
coupling is weak and the D3-brane exact solution can be trusted.
In the geometrical tachyon picture, this happens near the top of
the geometrical tachyon potential. As the geometrical tachyon
rolls down its potential, the inflation ends, and as the bulk
tachyon asymptotes to zero reaching the IR fixed point of the
background theory, the brane-universe enters a late accelerating
phase. This happens when the geometrical tachyon reaches the
bottom of the potential.

For the same choice of the parameters, the equation of state
parameter $w$, starts with the value $w=-1$ in the early
inflationary phase, and passing from positive values relaxes again
to $w=-1$, in the late accelerating phase. The whole evolution of
$w$ occurs without the need of introducing any dark matter or dark
energy. It is the dynamics of the bulk theory that drives the
whole evolution. Of course for a realistic cosmological evolution
of the brane-universe, matter fields have to be introduced on the
brane to account for the diversity of matter we observe in our
universe.

Type-0 is an interesting string background. We are however, far
away from understanding its full dynamics and its connection to
the boundary open string theory. Only very limited solutions are
known near the conformal fixed points of the theory. If for
example more general solutions of the Type-0 string background
were known with non-constant tachyon and dilaton fields, the
coupling of the bulk tachyon to the geometrical potential would
have been a function of the distance $r$ from the bulk Dp-branes.
The geometrical tachyon potential then, would get $r$-dependent
modifications effecting the dynamics of the geometrical tachyon as
it rolls down its potential and its cosmological evolution.

This work provides some evidence that closed tachyon condensation
may have some important consequences on the cosmological evolution
of the boundary theory. Of course we do not fully understand the
dynamics of the closed tachyon condensation and its connection to
the gravitational dynamics, but nevertheless we provided an
example in which the condensation of the bulk tachyon, except that
it stabilizes the bulk theory, it is responsible for the inflation
on the boundary theory.

More work is needed to understand the phenomenological aspects of
the inflation on the brane, like how long it lasts, what are the
scalar perturbations produced during inflation, what is the
mechanism for reheating. Can the condensation of the bulk tachyon
provide the energy needed for the reheating? However, there is a
drawback in these considerations, because the relation
(\ref{brel}) indicates that the kinetic energy of the geometrical
tachyon field is not small and it can not be ignored compared to
unity and hence slow-roll inflation can not be applied. This can
be understood because of the strong bulk effects of the bulk
tachyon condensation.

Our final remark conserns the consistency of our approach. Let us
recall what happens in the DGP model~\cite{Dvali:2000hr}. We have
a static brane in a time dependent five-dimensional pure
gravitational bulk. As it was showed in
\cite{Kofinas:2001es,Maeda:2003ar}, the introduction of a
four-dimensional scalar curvature term on the brane, simply
redefines the energy momentum tensor on the brane. If we go to the
picture in which the brane is moving and the bulk is static, it
can be shown~\cite{csaki,Cuadros-Melgar:2005ex}, that the $R$ term
has the same effect, it redefines the energy momentum tensor in
the junctions conditions which play the r$\hat{o}$le of the
equations of motion of the moving brane. In both pictures, there
is an effective four-dimensional Einstein equation which describes
consistently the cosmological evolution on the brane.

In the case of a Dp-brane moving in the background of other
Dp-branes the situation is much more
difficult~\cite{Kiritsis:2003mc}. The Dp-branes of the bulk are
solutions of a complicated usually string theory and the only
information we can get on the brane is only through the DBI
action. For this reason we use the rolling tachyon picture which
is described by a simple DBI action. In our case, the coupling of
gravity on the moving brane complicates further the picture. Our
basic assumption is the probe brane approximation. There is no
backreaction between the brane and the bulk. As the brane moves in
the gravitational field of the background branes it does not
disturb this background. This assumption led us to the Friedmann
equation (\ref{fried}).

\section{Acknowlegements}

Work supported  by (EPEAEK II)-Pythagoras (co-funded by the
European Social Fund and National Resources). We thank Malcolm
Fairbairn, Ansar Fayyazuddin, Fawad Hassan, Alex Kehagias, George
Kofinas, M. Sami, John Ward for very helpful discussions, comments
and remarks.

\end{document}